\documentclass{ws-procs975x65}

\usepackage{latexsym} 

\newcommand{\half}{\textstyle {1\over2} \displaystyle}    
\newcommand{\third}{\textstyle {1\over3} \displaystyle}   
\newcommand{\twoth}{\textstyle {2\over3} \displaystyle}   

\begin{document}

\title{Ultraviolet Divergences  and \\ Scale-Dependent Gravitational Couplings}

\author{Herbert W. Hamber
\footnote{Plenary Talk,  12-th Marcel Grossmann Conference on 
{\it Recent Developments in General Relativity, Astrophysics and Relativistic Field Theories}, 
UNESCO, Paris, July 12-18, 2009.}
}

\address{Department of Physics, University of California,\\
Irvine, CA 92617, USA\\
$^*$E-mail: HHamber@uci.edu}

\begin{abstract}

I review the field-theoretic renomalization group approach to quantum gravity, 
built around the existence of a non-trivial ultraviolet fixed point in four dimensions.
I discuss the implications of such a fixed point, found in three 
largely unrelated non-perturbative approaches,  and
how it relates to the vacuum state of quantum gravity, and specifically to the 
running of $G$.
One distinctive and well-known feature of the new fixed 
point is the emergence of a second genuinely non-perturbative scale, analogous 
to the scaling violation parameter in non-abelian gauge theories.
I argue that it is natural to identify such a scale with the small observed
cosmological constant, which in quantum gravity can arise as a non-perturbative
vacuum condensate.
I then show how the lattice cutoff theory of gravity can in principle provide
quantitative predictions on the running of
$G$, which can then be used to construct manifestly covariant effective field equations,
and from there estimate the size of non-local quantum corrections to the
standard GR framework.

\end{abstract}


\bodymatter

\vskip 20pt

\section{Perturbative Non-renormalizability and Feynman Path Integral}

\label{sec:one-loop}

In a quantum theory of gravity the coupling constant is dimensionful, $G \sim \mu^{2-d}$, 
and within the standard perturbative treatment of radiative corrections one expects trouble
in four dimensions, based on purely on dimensional grounds.
The divergent one loop corrections are proportional to
$G \Lambda^{d-2} $ where $\Lambda$ is the ultraviolet cutoff, which
then leads to a bad high momentum behavior, with an effective running 
Newton's constant
\begin{equation} 
G(k^2) \, / \,  G \,  \sim \, 1 + c_1 (d) \; G \, k^{d-2} + \, O(G^2) \; .
\label{eq:g-pert}
\end{equation}
A more general argument for perturbative non-renormalizability
starts by considering the gravitational action with scalar curvature term $R$,
which involves two derivatives of the metric.
Then the graviton propagator in momentum space goes like $1/k^2$,
and the vertex functions like $k^2$.
In $d$ dimensions each loop integral with involve a momentum
integration $d^d k$, so that the superficial degree of divergence ${\cal D}$
of a Feynman diagram with $L$ loops is given by
\begin{equation}
{\cal D} = 2 + (d-2) \, L \; ,
\end{equation}
independent of the number of external lines.
One therefore concludes that for $d>2$ the degree of
divergence for Einstein gravity  increases rapidly with loop order $L$, and
that the theory cannot be renormalized in naive perturbation theory.
A consequence of the lack of perturbative renormalizability is the fact
that new higher derivative counterterms arise to one-loop order  [1]
\begin{equation}
\Delta {\cal L}_g = { \sqrt{g} \over 8 \pi^2 (d-4) }
\left ( 
{1 \over 120 } R^2 + 
{ 7 \over 20 } R_{\mu\nu} R^{\mu\nu} 
\right ) \;  ,
\end{equation}
with even higher derivatives appearing at the next order.
One concludes that the standard approach based on a perturbative 
expansion of the pure Einstein theory in four dimensions is not 
convergent; in fact it is badly divergent.

A number of possible options have been proposed, the simplest of which
is to just add the above higher derivative terms to the original action.
The resulting extended theory is perturbatively renormalizabe to all orders, 
but suffers potentially from unitarity problems.
But these cannot be satisfactorily addressed in perturbation theory,
as the theory is now asymptotically free in the
higher derivative couplings and presumably exhibits a non-trivial vacuum.
Alternatively, the gravity divergences can be cancelled by adding 
new unobserved  massless particles
and invoking supersymmety; in fact it has been claimed recently that $N=8$ 
supergravity might not be just renormalizable, but indeed finite to some 
relatively high loop order.
The downside of this somewhat contrived approach is the artificial introduction
of a plethora of unobserved massless particles of spin $0,1/2,1,3/2$,
added to the original action in order to cancel  the gravitational ultraviolet divergences.
Finally, string theory solves the problem of ultraviolet divergences by postulating the existence
of fundamental extended objects, thus in part bypassing the requirement of supersymmetry
and providing a natural cutoff for gravity, related to a fundamental string scale [2].

Nevertheless one important point that cannot be overlooked is the fact that in other 
field theories,
which to some extent share with gravity the same set of ultraviolet problems
(the non-linear sigma model is the most notable one, and the best studied case), 
the analogous result of Eq.~(\ref{eq:g-pert}) is in fact known to be {\it incorrect}.
It appears as an artifact of naive perturbation theory, which in four dimensions 
does not converge, and seems to lead therefore to fundamentally incorrect answers.
The correct answer in these models is found instead either by expanding around the dimension
in which the theory {\it is} perturbatively renormalizable, or by solving it exactly in the large 
$N$ limit and then computing $1/N$ corrections, or by solving it numerically on a lattice.
The generic new feature in these models is the existence of a non-trivial fixed point
of the renormalization group [3-7], which is inaccessible by perturbation theory
in four dimensions, and radically alters the ultraviolet behavior of the theory.
\footnote{
After QED, the second most accurate prediction of quantum field theory to date
is for a perturbatively non-renormalizable theory, the $O(N)$ non-linear $\sigma$-model
in three dimensions, whose field theoretic treatment based on a non-trivial
fixed point of the renormalization group, either
on the lattice or in the continuum, eventually provides detailed predictions 
for scaling behavior and anomalous dimensions in the vicinity of the fixed point [5,8].
These have recently been verified experimentally to high accuracy in a sophisticated 
space shuttle experiment [9] for critical superfuid Helium, 
whose order parameter  corresponds to $N=2$ in the non-linear $\sigma$-model.
}
The key question for gravity is therefore: are the ultraviolet problems just 
an artifact of a naive application of perturbation theory in four dimensions, 
as clearly happens in other perturbatively non-renormalizable theories that
also contain dimensionful couplings in four dimensions?

In the following I will limit my discussion to the approach 
based on traditional quantum field theory methods and the renormalization
group, applied to the Einstein action with a cosmological term, 
an avenue which in the end is intimately tied with the existence of a non-trivial 
ultraviolet fixed point in $G$ in four dimensions.
The nature of such a fixed point was first discussed in detail by K. Wilson for scalar 
and fermionic theories [3], 
and the methods later applied to gravity in [6], where they were referred to as 
asymptotic safety.
As discussed above, it is fair to say that so far this is the only approach known to work
in other not perturbatively renormalizable theories.
  
If non-perturbative effects play an important role in quantum gravity, 
then one would expect the need for an improved formulation
of the quantum theory, which does not rely exclusively
on the framework of perturbation theory.
After all, the fluctuating quantum metric field $g_{\mu\nu}$ is dimensionless,
and carries therefore no natural scale.
For the somewhat simpler cases of a scalar field and non-Abelian gauge theories 
a consistent
non-perturbative formulation based on the Feynman path integral 
has been used for some time, and is by now well developed.
In a nutshell, the Feynman path integral formulation for quantum
gravitation can be expressed by the functional integral formula
\begin{equation}
Z = \int_ {\rm geometries }  e^{ \, { i \over \hbar} I_{\rm geometry} } \;\; .
\end{equation}
Furthermore a bit of thought reveals that for gravity, to all orders
in the weak field expansion, there is really no difference of
substance between the Lorentzian (or pseudo-Riemannian) and the
Euclidean (or Riemannian) formulation, which can be mapped into
each other by analytic continuation. In the following therefore
the Euclidean formulation will be assumed, unless stated otherwise.

In function space one needs a metric
before one can define a volume element.
Therefore, following DeWitt, one first defines an invariant norm for metric deformations  
\begin{equation}
\Vert \delta g \Vert^2 \, = \, 
\int d^d x \; \delta g_{\mu \nu}(x) \,
G^{\mu \nu, \alpha \beta} \bigl [ g(x) \bigr ] \,
\delta g_{\alpha \beta}(x) \;\; ,
\end{equation}
with the supermetric $G$ given by the ultra-local
expression
\begin{equation}
G^{\mu \nu, \alpha \beta} \bigl [ g(x) \bigr ] \, = \, 
\half \, \sqrt{g(x)} \, \left [ \,
g^{\mu \alpha}(x) \, g^{\nu \beta}(x) +
g^{\mu \beta}(x) \, g^{\nu \alpha}(x) - \lambda \,
g^{\mu \nu}(x) \, g^{\alpha \beta}(x) \, \right ]
\end{equation}
with $\lambda$ a real parameter, $\lambda \neq 2 / d $.
The DeWitt supermetric then defines a suitable functional volume element $\sqrt{G}$
in four dimensions,
\begin{equation}
\int [d \, g_{\mu\nu} ] \,  = \, 
\int \, \prod_x \, \prod_{\mu \ge \nu} \, d g_{\mu \nu} (x) \; .
\end{equation}
The Euclidean Feynman path integral for pure
Einstein gravity with a cosmological constant term is then written as
\begin{equation}
Z_{cont} \; = \; \int [ d \, g_{\mu\nu} ] \; \exp \, \left \{
- \lambda_0 \, \int d x \, \sqrt g \, + \, 
{ 1 \over 16 \pi G } \int d x \, \sqrt g \, R \, \right \} \;  .
\label{eq:zcont}
\end{equation}
An important aspect of this path integral is connected with
the global scaling properties of the action and the measure [10].
First one notices that in pure Einstein gravity with a bare cosmological 
constant term 
\begin{equation}
{\cal L} = - { 1 \over 16 \pi G_0} \, \sqrt{g} \, R \, + \, \lambda_0 \sqrt{g}
\end{equation}
one can rescale the metric by $ g_{\mu\nu} = \omega \, g_{\mu\nu}' $ with 
$\omega$ a constant, giving 
\begin{equation}
{\cal L} = - { 1 \over 16 \pi G_0} \, \omega^{d/2-1} \, \sqrt{g'} \, R' 
\, + \, \lambda_0 \, \omega^{d/2} \, \sqrt{g'}  \; .
\label{eq:rescale}
\end{equation}
This can then be interpreted as a rescaling of the two bare couplings
$ G_0 \rightarrow \omega^{-d/2+1} G_0  $, 
$\lambda_0 \rightarrow \lambda_0 \, \omega^{d/2} $,
leaving the dimensionless combination $G_0^d \lambda_0^{d-2}$ unchanged.
Therefore only the latter combination has physical meaning in pure gravity.
In particular, one can always choose the scale $\omega = \lambda_0^{-2/d}$,
so as to adjust the volume term to have a unit coefficient.
The implication of this last result is that pure gravity only contains one 
bare coupling $G_0$, 
besides the ultraviolet cutoff $\Lambda$ needed to regulate the quantum theory.

\vskip 20pt

\section{Gravity in $2+\epsilon$ Dimensions and Non-Trivial UV Fixed Point}

\label{sec:phaseseps}

In two dimensions the gravitational coupling becomes dimensionless, 
$G\sim \Lambda^{2-d}$,
and the theory appears perturbatively renormalizable.
In spite of the fact that the gravitational action reduces to a topological
invariant, it is meaningful to attempt to construct, 
in analogy to what was suggested originally by Wilson for scalar field theories, 
the theory perturbatively as a double series in $\epsilon=d-2$ and $G$.
The $2+\epsilon$ expansion for pure gravity then proceeds as follows [6,11].
First the gravitational part of the action
\begin{equation}
{\cal L} = - { \mu^\epsilon \over 16 \pi G} \, \sqrt{g} \, R \, + \, \lambda_0 \sqrt{g} \; , 
\label{eq:l-pure}
\end{equation}
with $G$ now dimensionless and $\mu$ an arbitrary momentum scale, 
is expanded by setting
\begin{equation}
g_{\mu\nu} \, \rightarrow \, \bar g_{\mu\nu} = g_{\mu\nu} \, + \, h_{\mu\nu} \; ,
\end{equation}
where $g_{\mu\nu}$ is the classical background field, and $h_{\mu\nu}$
a small quantum fluctuation.
The quantity ${\cal L}$ in Eq.~(\ref{eq:l-pure}) is naturally identified with
the bare Lagrangian, and the scale $\mu$ with a microscopic ultraviolet
cutoff $\Lambda$, corresponding to the inverse lattice spacing in the lattice formulation.
After the quantum fluctuations in $h_{\mu\nu}$ are integrated out and
the cosmological constant term gets rescaled, one obtains the following
result for the renormalization group beta function for $G$:
with $N_S$ scalar fields and $N_F$ Majorana fermion fields the result to two loops reads [12]
\begin{equation}
\mu { \partial \over \partial \mu } G  =
\beta (G) = \epsilon \, G \, - \, \beta_0 \, G^2 
\, - \, \beta_1 \, G^3 \, + \, O( G^4, G^3 \epsilon , G^2 \epsilon^2 ) \; ,
\label{eq:beta-twoloop}
\end{equation} 
with $\beta_0 = \twoth \, (25 - c) $ and
$ \beta_1 = { 20 \over 3 } \, (25 - c) $, and $c \equiv N_S+N_F/2$.
The physics of this result is contained in the fact that the gravitational 
$\beta$-function determines the scale dependence
of Newton's constant $G$, and has the shape shown in Fig. 1.

\begin{center}
\epsfxsize=7cm
\epsfbox{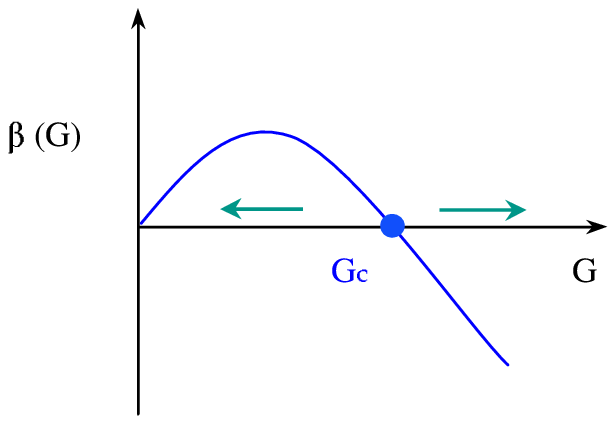}
\end{center}

\noindent{\small Figure 1.
Renormalization group $\beta$-function 
for gravity close to two dimensions. 
The arrows indicate the coupling constant flow towards 
increasingly larger distance scales.}
\medskip

\label{fig:beta-g-eps}

A closer examination of the result shows that
depending on whether one is on the right ($G>G_c$) or on the left 
($G<G_c$) of the non-trivial ultraviolet fixed point at
\begin{equation}
G_c = { d - 2 \over \beta_0 } + O((d-2)^2 )
\end{equation}
(with $G_c$ positive, provided one has $c<25$)
the coupling will either flow to increasingly larger values of $G$,
or flow towards the Gaussian fixed point at $G=0$, respectively.
Furthermore the running of $G$ as a function of the sliding momentum 
scale $\mu=k$ can be obtained by integrating 
Eq.~(\ref{eq:beta-twoloop}), and one has to lowest order
\begin{equation}
G(k^2) \; \simeq \; G_c \, \left [ 
\, 1 \, \pm \, c_0 \, \left ( {m^2 \over k^2 } \right )^{(d-2)/2}
\, + \, \dots \right ] \; ,
\label{eq:grun-cont1} 
\end{equation}
with $c_0$ a positive constant, and $m$ a new nonperturbative scale.
As in non-abelian gauge theories and QCD, this last quantity arises naturally as an 
integration constant of the renormalization group equations.
The choice of $+$ or $-$ sign is then determined from whether one is
to the left (+), or to right (-) of $G_c$, in which case
the effective $G(k^2)$ decreases or, respectively, increases as one flows away
from the ultraviolet fixed point towards lower momenta, or larger distances.
Physically therefore the two solutions represent a gravitational screening ($G<G_c$), and a
gravitational anti-screening ($G>G_c$) situation [10].
Finally, at energies sufficiently high to become comparable
to the ultraviolet cutoff, the gravitational coupling $G$ eventually flows towards the
ultraviolet fixed point 
$ G(k^2) \, \mathrel{\mathop\sim_{ k^2 \rightarrow \Lambda^2 }} \, G (\Lambda) $,
where $G(\Lambda)$ is the coupling at the cutoff scale $\Lambda$,
to be identified with the bare or lattice coupling. 

One message therefore is that the quantum corrections
involves a new physical, renormalization group invariant scale $\xi=1/m$,
which cannot be fixed perturbatively and whose size determines the 
scale for the new quantum effects.
In terms of the bare coupling $G(\Lambda)$, it is given by
\begin{equation}
\xi^{-1} = m = A_m \cdot \Lambda \, 
\exp \left ( { - \int^{G(\Lambda)} \, {d G' \over \beta (G') } }
\right ) \; .
\label{eq:m-cont}
\end{equation}
The constant $A_m$ on the r.h.s. of Eq.~(\ref{eq:m-cont})
cannot be determined by perturbation theory; 
it needs to be computed by non-perturbative lattice methods.

At the fixed point $G=G_c$ the theory is scale invariant by definition;
in statistical field theory language the fixed point corresponds to a phase transition.
In the vicinity of the fixed point one can write
\begin{equation}
\beta (G) \, \mathrel{\mathop\sim_{ G \rightarrow G_c }} \, 
\beta' (G_c) \, (G-G_c) \, + \, O ((G-G_c)^2 ) \; .
\label{eq:beta-lin-g}
\end{equation}
If one defines the exponent $\nu$ by $ \beta ' (G_c) \, = \, - 1/ \nu $,
then from Eq.~(\ref{eq:m-cont}) one has by integration 
\begin{equation}
m \, \mathrel{\mathop\sim_{G \rightarrow G_c }} \,
\Lambda \cdot A_m \, | \, G (\Lambda) - G_c |^{\nu} \;\; ,
\label{eq:m-cont1}
\end{equation}
with $\nu$ the correlation length exponent.
To two loops the results of [12] imply
\begin{equation}
\nu^{-1} = \epsilon \, + \, {15 \over 25 - c } \, \epsilon^2 \, + \, \dots
\label{eq:nueps}
\end{equation}
which gives, for pure gravity without matter ($c=0$) in four dimensions, to lowest order
the scaling exponent $\nu^{-1} = 2$, and $\nu^{-1} \approx 4.4 $ at the next order.
The key question raised by these $2 + \epsilon$ perturbative calculations is
therefore: what remains of the above phase transition in four dimensions,
how are the two phases of gravity characterized  non-perturbatively, 
and what is the value of the exponent $\nu$ determining
the running of $G$ in the vicinity of the fixed point 
in four dimensions ?
To answer this question in a controlled way would seem to require the 
introduction of a  non-perturbative regulator, based on the lattice formulation
(since no other reliable non-perturbative regulator for field theories is
known to date).

\vskip 20pt

\section{Lattice Regularized Quantum Gravity}

\label{sec:lattice}

On the lattice the infinite number of degrees of freedom in the continuum
is restricted, by considering Riemannian spaces described
by only a finite number of variables, to the geodesic distances between
neighboring points.
Such spaces are taken to be flat almost everywhere, and referred to as 
piecewise linear.
The elementary building blocks for $d$-dimensional space-time are then
simplices of dimension $d$.
A 0-simplex is a point, a 1-simplex is an edge, a 2-simplex is a triangle, a
3-simplex is a tetrahedron.
A $d$-simplex is a $d$-dimensional object with $d+1$ vertices and
$d(d+1)/2$ edges connecting them [13].

The geometry of the interior of a $d$-simplex is assumed to be flat,
and is therefore completely specified by the lengths of its $d(d+1)/2$ edges.
When focusing on one such $n$-simplex, it is convenient to label
the vertices by $0,1, 2, 3, \dots , n$ and denote the 
square edge lengths by $l _ {01}^2 = l _ {10}^2$, ... , $l _ {0n}^2 $.
The simplex can then be spanned by the set of $n$ vectors 
$e_1$, ... $e_n$ connecting the vertex $0$ to the 
other vertices.
To the remaining edges within the simplex one then 
assigns vectors $e_{ij} = e_i-e_j$ with $1 \le i < j \le n$.
Within each $n$-simplex one can define a metric 
$ g_{ij} (s) \; = \; e_i \cdot e_j  $, and then in terms of the edge lengths
$l_{ij} \, = \, | e_i - e_ j | $ the metric is given by
\begin{equation}
g_{ij} (s) \; = \; \half \, 
\left ( l_{0i}^2 + l_{0j}^2 - l_{ij}^2 \right ) \; .
\label{eq:latmet}
\end{equation}
The volume of a general $n$-simplex can be found by the $n$-dimensional
generalization of the well-known formula for a tetrahedron, namely
\begin{equation}
V_n (s) \; = \; {1 \over n ! }  \sqrt { \det  g_{ij} (s) } \; .
\label{eq:vol-met}
\end{equation}
In a piecewise linear space curvature is detected by going around
elementary loops which are dual to a ($d-2$)-dimensional subspace.
From the dihedral angles $\theta (s,h)$
associated with the faces of the simplices meeting
at a given hinge $h$ one computes the deficit angle $\delta (h)$,
defined as [13]
\begin{equation}
\delta (h) \; = \; 2 \pi \, - \, \sum_{ s \supset h } \; \theta (s,h) \; ,
\label{eq:deficit}
\end{equation}
where the sum extends over all simplices $s$ meeting on $h$.
It then follows that the deficit angle $\delta$ is a measure
of the local curvature at $h$.
By considering rotation matrices around a hinge one can obtain
an expression for the Riemann tensor at the hinge $h$ 
\begin{equation}
R_{\mu\nu\lambda\sigma} (h) \; = \; {\delta (h) \over A_C (h) } 
\, U_{\mu\nu} (h) \, U_{\lambda\sigma} (h)
\label{eq:riem-hinge}
\end{equation}
which is expected to be valid in the limit of small curvatures,
with $A_C (h) $ the area of the loop entangling the hinge $h$,
and $U$ a bivector describing the hinge's orientation.
From the expression for the Riemann tensor at a hinge
given in Eq.~(\ref{eq:riem-hinge}) one obtains by contraction the Ricci scalar
$ R (h) = 2 \, \delta (h) / A_C (h) $, and the
continuum expression $\sqrt{g} \, R$ is then obtained
by multiplication with the volume element $V (h) $ associated with
a hinge.
The curvature and cosmological constant terms then lead to the combined Regge lattice action
\begin{equation}
I_{\rm latt} (l^2) \; = \; \lambda_0 \sum_{\rm simplices \; s} \, V^{(d)}_s
\, - \, k \sum_{\rm hinges \; h} \,  \delta_h \, V^{(d-2)}_h \; .
\label{eq:latac}
\end{equation}
One key aspect of this formulation is the local gauge invariance of the 
lattice action, in analogy to the local gauge invariance of the Wilson
action for gauge theories.
Already on a flat 2-d lattice it is clear that one can move around a point on a surface, 
keeping all the neighbors fixed, without violating the triangle inequalities, and
leaving local curvature invariants unchanged.
In $d$ dimensions this transformation has $d$ parameters and is an exact
invariance of the action.
When space is slightly curved, piecewise linear diffeomorphisms
can still be defined as the
set of local motions of points that leave the local contribution to the action,
the measure and the lattice analogs of continuum curvature invariants unchanged.
In the limit when the number of edges becomes very large one expects the full continuum diffeomorphism group to be recovered [14].

In order to write down a lattice path integral, one needs, besides the action,
a functional measure.
As the edge lengths $l_{ij}$ play the role of the continuum metric
$g_{\mu\nu}(x)$, one expects the discrete measure to involve an
integration over the squared edge lengths.
After choosing coordinates along the edges emanating from a vertex,
the relation between metric perturbations and squared edge
length variations for a given simplex based at 0 in $d$ dimensions is
from Eq.~(\ref{eq:latmet})
\begin{equation}
\delta g_{ij} (l^2) \; = \; \half \;
( \delta l_{0i}^2 + \delta l_{0j}^2 - \delta l_{ij}^2 ) \; .
\label{eq:latmet1}
\end{equation}
For one $d$-dimensional simplex labeled by $s$
the integration over the metric is thus equivalent to an 
integration over the edge lengths, and one has the identity
\begin{equation}
\left ( {1 \over d ! } \sqrt { \det g_{ij}(s) } \right )
\prod_{ i \geq j } \, d g_{i j} (s) = 
{\textstyle \left ( - { 1 \over 2 } \right ) \displaystyle}^{ d(d-1) \over 2 }
\left [ V_d (l^2) \right ] \!
\prod_{ k = 1 }^{ d(d+1)/2 } \! \! dl_{k}^2 \; .
\label{eq:simpmeas}
\end{equation}
Indeed there are $d(d+1)/2$ edges for each
simplex, just as there are $d(d+1)/2$ independent components for the metric
tensor in $d$ dimensions.
In addition, a certain set of simplicial inequalities need to be imposed on the edge lengths.
These represent conditions conditions on the edge lengths $l_{ij}$ 
such that the sites $i$ can be considered as vertices of a
$d$-simplex embedded in flat $d$-dimensional Euclidean space.
After summing over all simplices one derives 
what is regarded as the lattice functional measure
representing the continuum DeWitt measure in four dimensions
\begin{equation}
\int [ d l^2] \; = \; \int_0^\infty \prod_{ ij } \, dl_{ij}^2 
\; \Theta [ l_{ij}^2 ] \;\; .
\label{eq:lattmeas}
\end{equation}
Here $ \Theta [l_{ij}^2] $ is
a (step) function of the edge lengths, with the property
that it is equal to one whenever the triangle inequalities and their
higher dimensional analogs are satisfied and zero otherwise.
The lattice action of Eq.~(\ref{eq:latac}) for pure four-dimensional Euclidean 
gravity then leads to the regularized lattice functional integral [7]
\begin{equation} 
Z_{latt} \; = \;  \int [ d \, l^2 ] \; \exp \left \{ 
- \lambda_0 \sum_h V_h \, + \, k \sum_h \delta_h A_h
\right \} \; ,
\label{eq:zlatt} 
\end{equation}
where, as customary, the lattice ultraviolet cutoff is set equal to one
(i.e. all length scales are measured in units of the lattice cutoff).
Furthermore, $\lambda_0$ sets the overall scale, and can therefore be set 
equal to one without any loss of generality, according to the scaling arguments
presented before.
The lattice partition function $Z_{latt}$ should then be compared to the
continuum Euclidean Feynman path integral for pure gravity of Eq.~(\ref{eq:zcont}).

In closing we note that what makes the Regge theory stand out compared to other
possible discretization  of gravity is the fact that it is the {\it only} lattice theory known 
to have the correct spectrum 
of continuous excitations in the weak field limit, i.e. transverse traceless modes, 
or, equivalently, helicity-two massless gravitons. 
Indeed one of the simplest possible problems that can be treated in lattice quantum
gravity is the analysis of small fluctuations about a fixed flat simplicial background.
In this case one finds that the lattice graviton propagator in a 
De Donder-like gauge is identical to the continuum expression [7].

\vskip 20 pt

\section{Strongly Coupled Gravity and Gravitational Wilson Loop}

\label{sec:strong}

As in non-abelian gauge theories, important information about the 
non-perturbative ground state of the theory can be gained by
considering the strong coupling limit.
In lattice gravity an expansion can be performed for large $G$ or small $k=1/8 \pi G$,
and the resulting series is in general expected to be useful up to some $k=k_c$,
where $k_c$ is the lattice critical point,
at which point the partition function $Z$ eventually develops a singularity.
One starts from the lattice regularized
path integral with action Eq.~(\ref{eq:latac}) and 
measure Eq.~(\ref{eq:lattmeas}).
The four-dimensional Euclidean lattice action usually contains
a cosmological constant and scalar curvature term as in Eq.~(\ref{eq:latac}),
\begin{equation} 
I_{latt} \; = \;  \lambda \, \sum_h V_h (l^2) \, - \, 
k \sum_h \delta_h (l^2 ) \, A_h (l^2) \; .
\label{eq:ilatt1} 
\end{equation}
The action only couples edges which belong either to
the same simplex or to a set of neighboring simplices, and can therefore
be considered {\it local} just like the continuum action; 
it leads to the lattice partition function defined in Eq.~(\ref{eq:zlatt}).
When doing an expansion in the kinetic term
proportional to $k$, it is convenient to include the
$\lambda$-term in the measure, and 
$Z_{latt}$ can then be expanded in powers of $k$,
\begin{equation} 
Z_{latt}(k) \; = \;  \int d \mu (l^2) \, \; e^{k \sum_h \delta_h \, A_h } 
\; = \;  \sum_{n=0}^{\infty} \, { 1 \over n!} \, k^n \, 
\int d \mu (l^2) \, \left ( \sum_h \delta_h \, A_h \right )^n \;\; .
\label{eq:zlatt-k}
\end{equation}
$Z (k) \, = \, \sum_{n=0}^{\infty} a_n \, k^n $
is analytic at $k=0$, so this expansion is well defined up to the
nearest singularity in the complex $k$ plane.

In the gravity case the analogs of the gauge variables of Yang-Mills
theories are given by the connections, so it is natural when computing the
gravitational Wilson loop [15] to look for a first order formulation of Regge gravity [16].
For each neighboring pair of simplices $s,s+1$ one 
can associate a Lorentz transformation $ R^{\mu}_{\;\; \nu} (s,s+1)$,
and one then might want to consider a near-planar closed loop $C$,
such as the one shown schematically in Fig.2.
Along a closed loop the overall rotation matrix is given by 
\begin{equation}
R^{\mu}_{\;\; \nu} (C) \; = \;
\Bigl [ \prod_{s \, \subset C}  R_{s,s+1} \Bigr ]^{\mu}_{\;\; \nu} 
\end{equation}

\begin{center}
\epsfxsize=7cm
\epsfbox{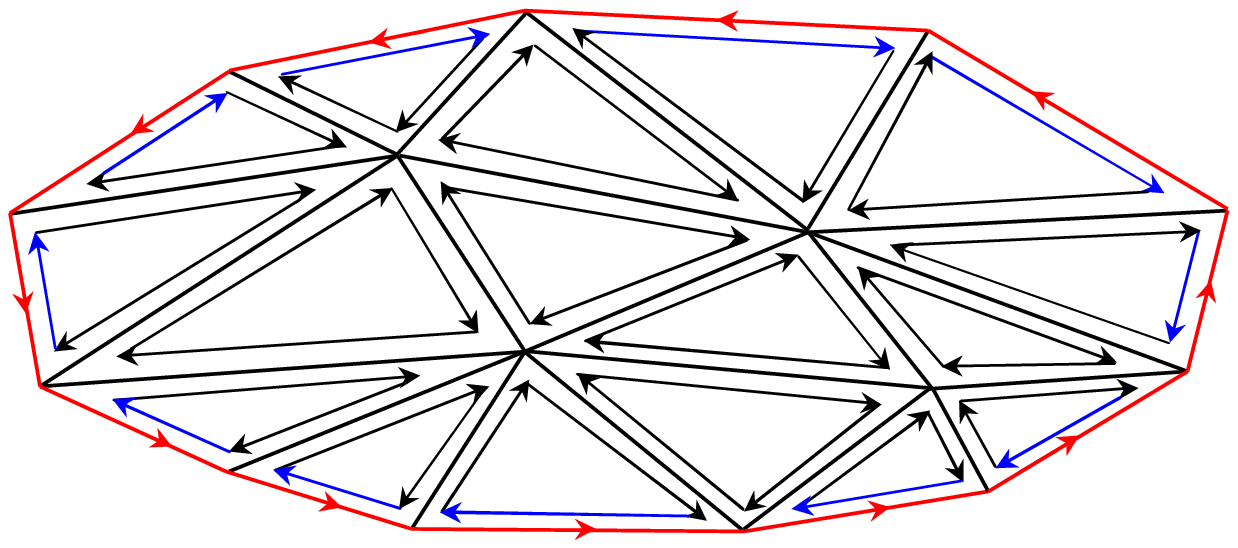}
\end{center}

\noindent{\small Figure 2.
Gravitational analog of the Wilson loop.
A vector is parallel-transported along the larger outer loop.
The enclosed minimal surface is tiled with parallel
transport polygons, here chosen to be triangles for illustrative
purposes.
For each link of the dual lattice, the elementary parallel transport
matrices ${\bf R}(s,s')$ are represented by arrows. }
\medskip

\label{fig:wilson}

In the quantum theory one is interested in averages
of the above product of rotations along a given path.
If the curvature of the manifold is small, then classically the expression
for the rotation matrix ${\bf R} (C )$ associated with a near-planar
loop can be re-written in terms of a surface
integral of the large-scale Riemann tensor, projected along the surface
area element bivector $A^{\alpha\beta} (C )$ associated with the orientation of
the loop,
\begin{equation}
R^{\mu}_{\;\; \nu} (C) \; \approx \; 
\Bigl [ \, e^{\half \int_S 
R^{\, \cdot}_{\;\; \cdot \, \alpha\beta} \, A^{\alpha\beta} ( C )} 
\Bigr ]^{\mu}_{\;\; \nu}  \;\; .
\label{eq:rotriem}
\end{equation}
Thus a direct calculation of the quantum Wilson loop could in principle
provide a way of determining the {\it effective} curvature on very
large distance scales, even in the case
where short distance fluctuations in the metric may be significant.

A detailed lattice calculation in the strong coupling limit then gives 
the following result.
First one defines the lattice Wilson loop as
\begin{equation}
 W(C)  \; = \; < \; Tr[(U_C \; + \; \epsilon \; I_4) \; 
R_1 \; R_2 \; ... \; ... \; R_n] \; > \; .
\label{eq:wloop-def2}
\end{equation}
where the $R_i$'s are the rotation matrices along the path and the factor 
$(U_C + \epsilon I_4)$ contains some average direction bivector
$U_C$ for the loop, which is assumed to be close to planar.
Then for sufficiently strong coupling one can show that one naturally
obtains an area law, which here we express as 
\begin{equation}
W ( C ) \, \simeq \, \exp ( - \, A_C / {\xi }^2 )
\label{eq:expdecay}
\end{equation}
where $\xi $ is the gravitational correlation length.
The appearance of $\xi$ follows from dimensional arguments, given that the
correlation length is the only relevant length scale in the vicinity of the
fixed point; the results can thus be considered analogous to the well-known 
behavior for the Wilson loop in non-abelian gauge theories [17].
In the actual calculation the rapid decay of the quantum gravitational Wilson 
loop as a function of the area can be seen as a general consequence of the
assumed disorder in the uncorrelated fluctuations of the parallel transport
matrices ${\bf R}(s,s')$ at large $G$.
A careful identification of (a suitable trace of) Eq.~(\ref{eq:rotriem}) 
with the expression in
Eq.~(\ref{eq:expdecay}), and in particular the comparison of the two 
area-dependent terms, then yields the following estimate
for the macroscopic, large scale, average curvature in the large  $G$ limit
\begin{equation}
{\bar R} \sim 1 / \xi^2 \; ,
\end{equation} 
where $\xi$ is the quantity in Eq.~(\ref{eq:m-cont}).
An equivalent way of phrasing the last result is the suggestion that $1 / \xi^2$,
where $\xi$ is the renormalization group invariant 
gravitational  correlation length of Eq.~(\ref{eq:m-cont}), should
be identified, up to a constant of proportionality of order one,
with the observationally determined, large scale cosmological constant $\lambda$.

\vskip 20 pt

\section{Nonperturbative Gravity}

\label{sec:numerical}

The exact evaluation of the lattice functional integral
for quantum gravity by numerical methods
allows one, in principle, to investigate a regime
which is generally inaccessible by perturbation theory: where
the coupling $G$ is strong and quantum fluctuations in the metric
are large.
The hope is, in the end, to make contact with the analytic
results obtained in the $2+\epsilon$
expansion, and determine which scenarios are physically
realized in the lattice regularized model.
The main question one would therefore like to answer is whether
there is any indication that the non-trivial ultraviolet fixed
point scenario is realized in the lattice theory, in four dimensions.
This would imply, 
as in the non-linear sigma model and similar models, the existence of at least
two physically distinct phases, and associated non-trivial scaling dimensions.
A clear physical characterization of the two gravitational phases would
also allow one, at least in principle, to decide which phase, 
if any, could be realized in nature.
As discussed below, the lattice continuum limit
is taken in the vicinity of the fixed point,
so close to it is perhaps the physically most relevant regime.

At the next level one would hope to be able to establish
a quantitative connection with the continuum
perturbative results, such
as the $2+\epsilon$ expansion discussed earlier.
Since the lattice cutoff and the method of dimensional regularization
cut the theory off in the ultraviolet in rather different
ways, one needs to compare universal quantities
which are {\it cutoff-independent}.
An example is the critical exponent $\nu$, as well as any other
non-trivial scaling dimension that might arise.
One should note that within the $2+\epsilon$ expansion only {\it one}
such exponent appears, to {\it all} orders in the loop
expansion, as $ \nu^{-1} = - \beta ' (G_c) $.
Therefore one central issue in the four-dimensional lattice regularized theory 
is the value of the universal scaling exponent $\nu$ [10,18] 
[in Eqs.~(\ref{eq:grun-cont1}),  (\ref{eq:m-cont1})  and (\ref{eq:nueps}) ] .

The starting point is again the lattice regularized
path integral with action as in Eq.~(\ref{eq:latac}) and
measure as in Eq.~(\ref{eq:lattmeas}).
Among the simplest quantum mechanical averages that one can compute 
is one associated with the local curvature,
\begin{equation}
{\cal R} (k) \; \sim \;
{ < \int d x \, \sqrt{ g } \, R(x) >
\over < \int d x \, \sqrt{ g } > } \;\; .
\end{equation}
But the curvature associated with this quantity is
one that would be detected when parallel-transporting 
vectors around very small infinitesimal loops.
Furthermore when computing correlations in quantum gravity new subtle 
issues arise, due to the fact that the physical distance between any two points $x$ and $y$
\begin{equation}
d(x,y \, \vert \, g) \; = \; \min_{\xi} \; \int_{\tau(x)}^{\tau(y)} d \tau \;
\sqrt{ \textstyle g_{\mu\nu} ( \xi )
{d \xi^{\mu} \over d \tau} {d \xi^{\nu} \over d \tau} \displaystyle } 
\end{equation}
is a fluctuating function of the background metric $g_{\mu\nu}(x)$.
Consequently physical correlations have to be defined at fixed
geodesic distance $d$, as in the following connected correlation
between observables $O$
\begin{equation}
< \int d x \, \int d y \, \sqrt{g} \, O(x) \; \sqrt{g} \, O(y) \;
\delta ( | x - y | - d ) >_c \; .
\end{equation}
Based on general arguments 
one expects such correlations to either follow a power law decay
at short distances, or an exponential decay characterized by a correlation
 length $\xi$ at larger distances
\begin{equation}
< \sqrt{g} \; O(x) \; \sqrt{g} \; O(y) \; \delta ( | x - y | -d ) >_c
\; \; \mathrel{\mathop\sim_{d \; \gg \; \xi }} \;\;
e^{-d / \xi } \;\;\;\; .
\label{eq:rr-exp}
\end{equation}
In practice such correlations at fixed geodesic distance
are difficult to compute numerically, and therefore not the best route
to study the critical properties of the theory.
But scaling arguments allow one to determine the scaling
behavior of correlation functions from critical exponents
characterizing the singular behavior of the free energy
$F (k) = - (1/V) \ln Z $ and of various local averages in the vicinity of the critical point.
In general a divergence of the correlation length $\xi$
\begin{equation}
\xi (k) \; \mathrel{\mathop\sim_{ k \rightarrow k_c}} \; A_\xi \;
| k_c - k | ^{ -\nu }
\label{eq:xi-k}
\end{equation}
signals the presence of a phase transition, and leads to the appearance
of a non-analyticty in the free energy $F(k)$.
One way to determine $\nu$ is from the curvature fluctuation, for which
one can show
\begin{equation}
\chi_{\cal R} (k) \; \mathrel{\mathop\sim_{ k \rightarrow k_c}} \;
A_{ \chi_{\cal R} } \; | k_c - k | ^{ -(2- d \nu) } \;\;\;\; .
\label{eq:chising}
\end{equation}
From such averages and fluctutations one can therefore, in principle, extract
the correlation length exponent $\nu$ of 
Eq.~(\ref{eq:xi-k}), without having to compute an invariant correlation
function at fixed geodesic distance.

In general for the measure in Eq.~(\ref{eq:lattmeas}) 
one finds a well behaved ground state only for $k < k_c $ [10].
The system then resides in the `smooth' phase, with an effective 
dimensionality close to four.
On the other hand, for $k > k_c$ the curvature becomes very large
and the lattice collapses locally into degenerate configurations
with very long, elongated simplices.
This last phenomenon is usually interpreted as a lattice remnant of the conformal
mode instability of Euclidean gravity.

There are a number of ways by which the critical exponents can
be determined to some accuracy from numerical simulations,and it is beyond
the scope of this short review to go into more details;  one obtains eventually
in $d=4$ $ k_c \simeq  0.0636 $ and $ \nu \simeq 0.335 $, which suggests 
\begin{equation}
\nu = 1/3
\end{equation}
 for pure quantum gravity in four dimensions [18].
Note that at the critical point the gravitational coupling
is not weak, since $G_c \approx 0.626 $ in units of the ultraviolet
cutoff. 
Fig. 3 shows a comparison of the critical exponent $\nu$ obtained
by three independent methods, namely the original lattice result in $d=2,3,4$ [18], 
the recent $2+\epsilon$ expansion 
(to one and two loops) [12], 
and the even more recent renormalization group truncation method [19-21].

\begin{center}
\epsfxsize=8cm
\epsfbox{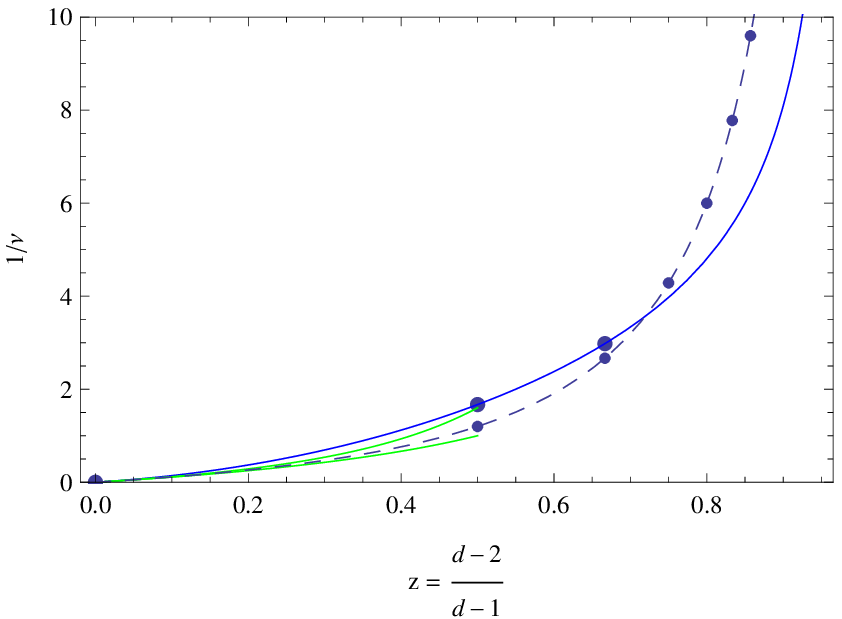}
\end{center}

\noindent{\small Figure 3.
Universal renormalization group scaling exponent $1/\nu$ of
Eq.~(\ref{eq:beta-lin-g}), computed in the
lattice theory of gravity (large dots for $d=2$, $d=3$ and $d=4$, 
and continous interpolating line) [18].
For comparison the $2+\epsilon$ result is shown
 (two lower curves, to one (lower) and two 
(upper) loops) [12],  as well as the recent Einstein-Hilbert truncation result
(smaller dots and connecting dashed line) [21].
The abscissa is, for convenience, a variable related to the space-time
dimension $d$ through $z=(d-2)/(d-1)$, which maps  
$d=\infty$ (where it is known that $\nu=0$) to $z=1$.
Note that for a scalar theory one has $1/ \nu=2$ for $d \geq 4$ [3]. }
\medskip

\label{fig:exp}

\section{Renormalization Group, Lattice Continuum Limit and the Running of $G$}
\label{sec:contlim}

The lattice theory points to the existence
of a phase transition in pure quantum gravity, with a divergent
correlation length in the vicinity of the critical point,
as in Eq.~(\ref{eq:xi-k}), which can be re-written
in terms of the inverse correlation length $m \equiv 1 / \xi$
\begin{equation}
\xi^{-1} \, = \, m \, = \, \Lambda \, A_m \, | \, k_c \, - \, k \, |^{ \nu } \;\; .
\label{eq:m-latt}
\end{equation}
In the above expression the correct dimensions have been restored,
 by inserting explicitly on the r.h.s. the ultraviolet
cutoff $\Lambda$.
Here $k$ and $k_c$ are dimensionless quantities, 
corresponding to bare microscopic couplings at the
cutoff scale, $k \equiv k (\Lambda) \equiv 1/ 8 \pi G(\Lambda) $.
$A_m$ is related to $A_\xi$ in Eq.~(\ref{eq:xi-k}) by $A_m = A_\xi^{-1}$.
It is worth pointing out that the above expression for $m (k) $ is identical in structure
to the $2+\epsilon$ result for continuum gravity, Eq.~(\ref{eq:m-cont1}).

Then the lattice continuum limit corresponds to a large cutoff limit,
taken at {\it fixed} $m$ or $\xi$,
\begin{equation}
\Lambda \rightarrow \infty \; , 
\;\;\;\; k \rightarrow k_c \; ,
\;\;\;\; m \; {\rm fixed} \; ,
\end{equation}
which shows that the continuum limit is in fact reached in the
vicinity of the ultraviolet fixed point,  $ k \rightarrow k_c$. 
In practice, since the cutoff ultimately determines
the physical value of Newton's constant $G$, the cutoff $\Lambda$
cannot be taken to $\infty$, and it persists as a fundamental scale
in the theory.
A very large value will suffice, $\Lambda^{-1} \sim 10^{-33} cm$,
for which it will still be true that $ \xi \gg \Lambda^{-1}$, which
is all that is required for the continuum limit.

In order to discuss the renormalization group behavior
of the coupling in the lattice theory it is convenient
to re-write the result of Eq.~(\ref{eq:m-latt})
directly in terms of Newton's constant $G$ as
\begin{equation}
m \, = \,  \Lambda \, \left ( { 1 \over c_0 } \right )^\nu \, 
\left [ { G ( \Lambda ) \over G_c } - 1 \right ]^ \nu 
\; ,
\label{eq:m-latt1}
\end{equation}
with the dimensionless constant $c_0$ related to
$A_m$ by $A_m = 1 / (c_0 k_c)^\nu $.
The above expression only involves the dimensionless
ratio $G(\Lambda)/G_c$, which is the only relevant
quantity here.
From the knowledge of the dimensionless constant 
$A_m$ in Eq.~(\ref{eq:m-latt})
one can estimate from first principles the value of $c_0$ in 
Eq.~(\ref{eq:m-latt1}) and later in Eq.~(\ref{eq:grun-k}).
Lattice results for the correlation functions at fixed geodesic distance
give a value for $A_m \approx 0.72 $ with a significant uncertainty,
which, when combined with the values $k_c \simeq 0.0636 $ and 
$ \nu \simeq 0.335$ given above, gives 
$c_0 = 1 /( k_c \, A_m^{1/\nu}) \simeq 42$.
Then the renormalization group invariance of 
$m=\xi^{-1}$ requires that the running gravitational
coupling $G(\mu)$ varies in the vicinity of the
fixed point in accordance with the above equation, with
$\Lambda \rightarrow \mu$, where $\mu$ is an arbitrary momentum scale,
\begin{equation}
m \, = \,  \mu \, \left ( { 1 \over c_0 } \right )^\nu \,
\left [ { G ( \mu ) \over G_c } - 1 \right ]^ \nu \; .
\label{eq:m-mu}
\end{equation}
The latter is equivalent to the renormalization group
invariance requirement
\begin{equation}
\mu \, { d \over d \, \mu } \, m ( \mu , G( \mu ) ) \, = \, 0
\label{eq:m-rg-latt}
\end{equation}
provided $G(\mu)$ is varied in a specific way.
Thus Eq.~(\ref{eq:m-rg-latt}) can be used to obtain
a Callan-Symanzik $\beta$-function for the coupling 
$G(\mu)$ in units of the ultraviolet cutoff,
\begin{equation}
\mu \, { \partial \over \partial \, \mu } \, G ( \mu ) \; = \; 
\beta ( G ( \mu ) ) \;\; ,
\label{eq:beta-g-mu}
\end{equation}
with $\beta (G)$ given in the vicinity of the non-trivial fixed point,
from Eq.~(\ref{eq:m-mu}), by
$ \beta (G ) \mathrel{\mathop\sim_{ G \rightarrow G_c}}  - \, { 1 \over \nu } \, ( G- G_c ) $.
Or one can obtain the scale dependence of the gravitational
coupling directly from Eq.~(\ref{eq:m-mu}),  which then gives
\begin{equation}
G( \mu ) \; = \;  
G_c \left [ 1 \, + \, c_0 (m^2 / \mu^2 )^{1 / 2 \nu } \, + \,
O ( \, ( m^2 / k^2 )^{1 \over \nu} ) \,  \right ]
\label{eq:grun-k} 
\end{equation}
in the physical anti-screening phase.
Again, this last expression can be compared directly to the lowest order 
$2+\epsilon$ result of Eq.~(\ref{eq:grun-cont1}).
The physical dimensions of $G$ can be restored, by multiplying the
above expression on both sides by the ultraviolet cutoff $\Lambda$,
if one so desires.
Physically the above lattice result implies
anti-screening: the gravitational coupling $G$ increases slowly with distance.

\vskip 20 pt

\section{Curvature Scales and Gravitational Condensate}

\label{sec:curvature}

The renormalization group running of $G (\mu)$ in Eq.~(\ref{eq:grun-k})
involves an invariant scale $\xi=1/m$.
At first it would seem that such a scale could take any value, including
a very small one based on the naive estimate $\xi \sim l_P$ - which would
then preclude any observable quantum effects in the foreseeable future.
But the results from the gravitational Wilson loop at strong coupling
would suggest otherwise, namely that the non-perturbative scale $\xi$
is in fact related to macroscopic {\it curvature}.
From astrophysical observation the average curvature is very small [22], 
so one would conclude that $\xi$ has to be very
large and possibly macroscopic,
\begin{equation}
\lambda_{obs} \; \simeq \; { 1 \over \xi^2 } 
\label{eq:xi_lambda}
\end{equation} 
with $\lambda_{obs}$ the observed small but non-vanishing scaled cosmological constant.
A further indication that the identification of the observed cosmological
constant with a mass-like - and therefore renormalization group invariant - term 
might make sense beyond the weak field limit can be seen, for example,
by comparing the structure of the three classical field equations
\begin{eqnarray}
R_{\mu\nu} \, - \, \half \, g_{\mu\nu} \, R \, + \, \lambda \, g_{\mu\nu} \; 
& = & \; 8 \pi G  \, T_{\mu\nu}
\nonumber \\
\partial^{\mu} F_{\mu\nu} \, + \, \mu^2 \, A_\nu \, 
& = & \; 4 \pi e \, j_{\nu} 
\nonumber \\
\partial^{\mu} \partial_{\mu} \, \phi \, + \, m^2 \, \phi \; 
& = & \; {g \over 3!} \, \phi^3
\label{eq:masses}
\end{eqnarray}
for gravity, QED (made massive via the Higgs mechanism) and a self-interacting scalar
field, respectively.
Nevertheless it seems so far that the strongest 
argument suggesting the identification of the scale $\xi$ with  $\lambda$ is derived
from the calculation of the gravitational Wilson loop at strong [16].

This relationship, taken at face value, implies a very large, cosmological value
for $\xi \sim 10^{28} cm$, given the present bounds on $\lambda_{phys}$.
Thus a set of modified Einstein equations, incorporating the
quantum running of $G$, would read
\begin{equation}
R_{\mu\nu} \, - \, \half \, g_{\mu\nu} \, R \, + \, \lambda \, g_{\mu\nu}
\; = \; 8 \pi \, G(\mu)  \, T_{\mu\nu}
\label{eq:field0}
\end{equation}
with $\lambda \simeq 1 / \xi^2  $, and $G(\mu)$ on the r.h.s. 
scale-dependent, in accordance with
Eq.~ (\ref{eq:grun-k}).
The precise meaning of $G(\mu)$ in a covariant framework
is given below.

\vskip 20 pt

\section{Effective Covariant Field Equations }

\label{sec:effective}

The result of Eq.~(\ref{eq:grun-k}) implies a 
running gravitational coupling in the vicinity of the
ultraviolet fixed point, with $m=1/\xi$, $c_0 > 0$ and $\nu \simeq 1/3$.
Since $\xi$ is expected to be very large,
the quantity $G_c$ in the above expression should now
be identified with the laboratory scale value 
$ \sqrt{G_c} \sim 1.6 \times 10^{-33} cm$.
The effective interaction in real space is then obtained by Fourier transform,
but since the above expression is singular as $k^2 \rightarrow 0$,
the infrared divergence needs to be regulated, which can be achieved
by utilizing as the lower limit of momentum integration
$m=1/\xi$.
A properly infrared regulated version of the above would read
\begin{equation}
G(k^2) \; \simeq \; G_c \left [ \; 1 \, 
+ \, c_0 \left ( { m^2 \over k^2 \, + \, m^2 } \right )^{1 \over 2 \nu} \, 
+ \, \dots \; \right ] \; .
\label{eq:grun-k-reg}
\end{equation}
Then at very large distances $r \gg \xi$  the gravitational coupling is
expected to approach the finite value $G_\infty = ( 1 + c_0 + \dots ) \, G_c $.

The first step in analyzing the consequences of a running of $G$
is to re-write the expression for $G(k^2)$ in a coordinate-independent
way, for example by the use of a non-local Vilkovisky-type effective action.
Since in going from momentum to position space one 
usually employs $k^2 \rightarrow - \Box$,
to obtain a quantum-mechanical running of the gravitational
coupling one has to make the replacement
$ G  \;\; \rightarrow \;\; G( \Box ) $.
Therefore from Eq.~(\ref{eq:grun-k}) one obtains
\begin{equation}
G( \Box ) \, = \, G_c \left [ \; 1 \, 
+ \, c_0 \left ( { 1 \over \xi^2 \Box  } \right )^{1 / 2 \nu} \, 
+ \, \dots \, \right ] \; ,
\label{eq:grun-box}
\end{equation}
and the running of $G$ is expected to lead to
a non-local gravitational action, for example of the form
\begin{equation}
I_{eff} \; = \; { 1 \over 16 \pi G } \int dx \sqrt{g} \,
\left [ 1 \, - \, c_0 \, 
\left ( {1 \over \xi^2 \Box } \right )^{ 1 / 2 \nu} \, 
+ \dots \right ] R \; \; .
\label{eq:ieff_sr}
\end{equation}
Due to the appearance of a fractional exponent,
the covariant operator appearing in the above expression
has to be suitably defined by analytic continuation. 
The latter can be done, for example, by computing $\Box^n$ for positive
integer $n$ and then analytically continuing to $n \rightarrow -1/2\nu$.

Alternatively, had one not considered the action of Eq.~(\ref{eq:ieff_sr})
as a starting point for
constructing the effective theory, one would naturally be led 
(as suggested by Eq.~(\ref{eq:grun-box}))
to consider instead the following effective field equation
\begin{equation}
R_{\mu\nu} \, - \, \half \, g_{\mu\nu} \, R \, + \, \lambda \, g_{\mu\nu}
\; = \; 8 \pi \, G  ( \Box )  \, T_{\mu\nu} \; ,
\label{eq:field1}
\end{equation}
the argument again being the replacement 
$G \, \rightarrow \, G(\Box) $ in the classical Einstein field equations.
Being manifestly covariant, these expressions at least satisfy some
of the requirements for a set of consistent field equations
incorporating the running of $G$.

The effective field equations of Eq.~(\ref{eq:field1}) can in fact be re-cast in a form
very similar to the classical field equations but
with a $ {\tilde T_{\mu\nu}} \, = \, \left [ G  ( \Box )  / G   \right ] \, T_{\mu\nu}$
defined as an effective, or gravitationally dressed, energy-momentum tensor.
Just like the ordinary Einstein gravity case,
in general ${\tilde T_{\mu\nu}}$ might not be covariantly conserved a priori,
$\nabla^\mu \, {\tilde T_{\mu\nu}} \, \neq \, 0 $, but ultimately the
consistency of the effective field equations demands that it
be exactly conserved, in consideration of the Bianchi identity satisfied
by the Riemann tensor.
In this picture therefore the running of $G$ can be viewed as contributing to a sort of 
"vacuum fluid", introduced in order to to account for the gravitational vacuum polarization
contribution.

\vskip 20 pt

\section{Static Isotropic Solutions}

\label{sec:static}

One can show that the quantum correction due
to the running of $G$ can be described - at least in the non-relativistic
limit of Eq.~(\ref{eq:grun-k-reg}) when applied to Poisson's equation -
in terms of a vacuum energy density $\rho_m(r)$, distributed around
the static source of strength $M$ in accordance with
\begin{equation}
\rho_m (r) \; = \; { 1 \over 8 \pi } \, c_{\nu} \, c_0 \, M \, m^3 \,
( m \, r )^{ - {1 \over 2} (3 - {1 \over \nu}) }  
\, K_{ {1 \over 2} ( 3 - {1 \over \nu} ) } ( m \, r ) 
\label{eq:rho_vac}
\end{equation}
and with $ c_{\nu} \; \equiv \;  2^{ {1 \over 2} (5 - {1 \over \nu}) } /
\sqrt{\pi} \, \Gamma( {1 \over 2 \, \nu} ) $,
and
\begin{equation}
4 \, \pi \, \int_0^\infty \, r^2 \, d r \, \rho_m (r) \; = \;  c_0 \, M  \;\; .
\label{eq:rho_vac2}
\end{equation}
More generally in the fully relativistic case, 
after solving the covariant effective field equations
with  $G(\Box)$ for $\nu =1/3$ one finds in Schwarzschild coordinates, and
 in the limit $r \gg 2 M G $,
\begin{equation}
A^{-1} (r) \; = \; = \; B (r) \; = \; 1 \, - { 2 \, M \, G \over r } \, + \, 
{4 \, c_0 \, M \, G \, m^3 \over 3 \, \pi } \, r^2 \, \ln \, ( m \, r ) 
\, + \, \dots
\label{eq:a_small_r3}
\end{equation}
The last expressions for $A(r)$ and $B(r)$ are therefore consistent with a
gradual slow increase  in $G$ with distance, in accordance with the formula
\begin{equation}
G \; \rightarrow \; G(r) \; = \; 
G \, \left ( 1 \, + \, 
{ c_0 \over 3 \, \pi } \, m^3 \, r^3 \, \ln \, { 1 \over  m^2 \, r^2 }  
\, + \, \dots
\right )
\label{eq:g_small_r3}
\end{equation}
in the regime $r \gg 2 \, M \, G$.
The last result is in some ways reminiscent of the QED small-$r$ result
\begin{equation}
Q \; \rightarrow \; Q(r) \; = \; Q \, \left ( 1 \, + \, 
{\alpha \over 3 \, \pi } \, \ln { 1 \over m^2 \, r^2 } \, + \, \dots
\right )
\label{eq:qed_s}
\end{equation}
In the gravity case, the correction vanishes as $r$ goes to zero: in this
limit one is probing the bare mass, unobstructed by its virtual graviton cloud.
In some ways the running $G$ term acts as a local cosmological
constant term, for which the
$r$ dependence of the vacuum solution for small $r$ is fixed by the nature
of the Schwarzschild solution with a cosmological constant term.
One could in fact wonder what these solutions might look like in $d$
dimensions, and after some straightforward calculations one finds
that in $d \ge 4 $ space-time dimensions a solution to the effective field equations
can only be found if  in Eq.~(\ref{eq:grun-box}) $\nu=1/(d-1)$ exactly [23].

\vskip 20 pt

\section{Cosmological Solutions}

\label{sec:cosm}
A scale dependent Newton's constant is expected to lead to small modifications
of the standard cosmological solutions to the Einstein field equations.
Here I will summarize what modifications are
expected from the effective field equations on the basis of $G(\Box)$,
as given in Eq.~(\ref{eq:grun-box}), which itself originates in
Eqs.~(\ref{eq:grun-k-reg}) and (\ref{eq:grun-k}).
The starting point are the quantum effective field equations
of Eq.~(\ref{eq:field1}), 
with $G(\Box)$ defined in Eq.~(\ref{eq:grun-box}).
In the Friedmann-Robertson-Walker (FRW) framework these are
applied to the standard homogeneous isotropic metric
\begin{equation}
d \tau^2 \; = \;  dt^2 - a^2(t) \left \{ { dr^2 \over 1 - k\,r^2 } 
+ r^2 \, \left( d\theta^2 + \sin^2 \theta \, d\varphi^2 \right)  \right \} \; .
\end{equation}
It should be noted that there are in fact {\it two} related quantum contributions to the
effective covariant field equations. 
The first one arises because of the presence of a non-vanishing 
cosmological constant $\lambda \simeq 1 / \xi^2 $, caused by the
non-perturbative vacuum condensate of Eq.~(\ref{eq:xi_lambda}).
As in the case of standard FRW cosmology, this is expected to be 
the dominant contributions at large times $t$, and gives an exponential
(for $\lambda>0$), or cyclic (for $\lambda < 0$) expansion of the scale factor.
The second contribution arises because of the explicit running of $G (\Box)$ in the 
effective field equations.
The next step is therefore a systematic examination of the nature of
the solutions to the full effective field equations,
with $G ( \Box )$ involving the relevant covariant d'Alembertian operator
\begin{equation}
\Box \; = \; g^{\mu\nu} \nabla_\mu \nabla_\nu 
\end{equation}
acting on second rank tensors as in the case of $T_{\mu\nu}$,
\begin{eqnarray}
\nabla_{\nu} T_{\alpha\beta} \, = \, \partial_\nu T_{\alpha\beta} 
- \Gamma_{\alpha\nu}^{\lambda} T_{\lambda\beta} 
- \Gamma_{\beta\nu}^{\lambda} T_{\alpha\lambda} \, \equiv \, I_{\nu\alpha\beta}
\nonumber
\end{eqnarray}
\begin{equation} 
\nabla_{\mu} \left( \nabla_{\nu} T_{\alpha\beta} \right)
= \, \partial_\mu I_{\nu\alpha\beta} 
- \Gamma_{\nu\mu}^{\lambda} I_{\lambda\alpha\beta} 
- \Gamma_{\alpha\mu}^{\lambda} I_{\nu\lambda\beta} 
- \Gamma_{\beta\mu}^{\lambda} I_{\nu\alpha\lambda}  \; .
\end{equation}
To start the process, one assumes for example that $T_{\mu\nu}$ has a perfect fluid form, 
for which one obtains  the action of $\Box^n$ on $T_{\mu\nu}$,  and 
then analytically continues to negative fractional values of $n = -1/2 \nu $.
Even in the simplest case, with  $G(\Box)$ acting on a {\it scalar} such as the trace
of the energy-momentum tensor $T^\lambda_{\; \lambda}$, one finds for the choice
$ \rho(t) = \rho_0 \, t^{\beta} $ and $ a(t) = r_0 \, t^{\alpha} $
the rather unwieldy expression
\begin{equation}
\Box^n \left[ - \rho (t) \right] \rightarrow 4^n \, 
\left( - 1 \right)^{n+1} { \Gamma \left( {\beta \over 2} + 1 \right) \, 
\Gamma \left( {\beta + 3 \, \alpha + 1 \over 2} \right) \over 
\Gamma \left( {\beta \over 2} + 1 - n \right) \, 
\Gamma \left( {\beta + 3\, \alpha + 1 \over 2} - n \right) } \, \rho_0 \, t^{\beta - 2 n} \; ,
\end{equation}
with integer $n$ then analytically continued to $n \rightarrow - {1 \over 2 \, \nu}$, with
$\nu=1/3$.

A more general calculation shows that a non-vanishing pressure contribution is generated in 
the effective field equations, even if one initially assumes a pressureless fluid, $p(t)=0$.
After a somewhat lengthy computation one obtains for a universe filled with non-relativistic 
matter ($p$=0) the following set of effective Friedmann equations, namely
\begin{eqnarray}
{ k \over a^2 (t) } \, + \,
{ \dot{a}^2 (t) \over a^2 (t) }  
& \; = \; & { 8 \pi G(t) \over 3 } \, \rho (t) \, + \, { 1 \over 3 \, \xi^2 }
\nonumber \\
& \; = \; & { 8 \pi G \over 3 } \, \left [ \,
1 \, + \, c_t \, ( t / \xi )^{1 / \nu} \, + \, \dots \, \right ]  \, \rho (t)
\, + \, \third \, \lambda 
\label{eq:fried_tt}
\end{eqnarray}
for the $tt$ field equation, and
\begin{eqnarray}
{ k \over a^2 (t) } \, + \, { \dot{a}^2 (t) \over a^2 (t) }
\, + \, { 2 \, \ddot{a}(t) \over a(t) } 
& \; = \; & - \, { 8 \pi G \over 3 } \, \left [ \, c_t \, ( t / \xi )^{1 / \nu} 
\, + \, \dots \, \right ] \, \rho (t) 
\, + \, \lambda
\label{eq:fried_rr}
\end{eqnarray}
for the $rr$ field equation.
In the above equations the running of $G$ appropriate for 
the Robertson-Walker metric is given by
\begin{equation}
G(t) \; = \; G \, \left [ \; 1 \, + \, c_t \, 
\left ( { t \over \xi } \right )^{1 / \nu} \, + \, \dots \, \right ] \; ,
\label{eq:grun_frw}
\end{equation}
with $c_t$ of the same order as $c_0$ of Eq.~(\ref{eq:grun-k}).
Note that it is the running of $G$ that induces an effective pressure term in the second 
($rr$) equation, corresponding to the presence of a relativistic fluid
arising from the vacuum polarization contribution.
The second important feature of the new equations is an additional power-law
acceleration contribution, in addition to the standard one due to $\lambda$.

\vskip 20 pt

\section{Quantum Gravity and Cosmological Density Perturbations}

\label{sec:densa}

Besides the cosmic scale factor evolution and the static isotropic solutions
just discussed, the running of $G(\Box)$ also affects
the nature of matter density perturbations on very large scales.
In discussing these effects, it is customary to introduce a perturbed metric of
the form
\begin{equation}
{d\tau}^2 = {dt}^2 - a^2 \left( \delta_{ij} + h_{ij} \right) dx^i dx^j
\label{eq:pert-metric}
\end{equation}
with $a(t)$ the unperturbed scale factor and $ h_{ij} (\vec{x},t)$ a small
metric perturbation.
The next step is to determine the effects of the running of $G$ on the relevant
matter and metric perturbations, again by the use of the modified field equations.
For sufficiently small perturbations, one can expand $G(\Box)$ appearing in the 
effective covariant field equations in powers of the metric perturbation $h_{ij} $ as
\begin{equation}
G(\Box) \; = \; G_0 \, 
\left[ 
1 + \, { c_0 \over \xi^{1 / \nu} } \,  \left( { 1 \over \Box^{(0)} } \right)^{1 / 2  \nu} 
\, \left( 1 - {1 \over 2  \nu} \, {1 \over \Box^{(0)}} \, \Box^{(1)} (h) + \dots \right) 
\right] 
\label{eq:gbox_h}
\end{equation}
It is also customary to expand the density, pressure and metric trace perturbation
modes in spatial Fourier components
\begin{equation}
\delta \rho (\vec{x},t) = \delta \rho (t) \, e^{i \, \vec{k}\,\cdot \, \vec{x}}
\;\;\;\;
\delta p (\vec{x},t) = \delta p (t) \, e^{i \, \vec{k}\,\cdot \, \vec{x}}
\;\;\;\;
h (\vec{x},t) = h(t)\, e^{i \, \vec{k}\,\cdot \, \vec{x}} \; .
\end{equation}
Normally the Einstein field equations 
$ R_{\mu\nu} - {1 \over 2} g_{\mu \nu} R + \lambda \, g_{\mu \nu} = 8 \pi G \, T_{\mu \nu} \, $
are given to first order in the small perturbations by
\begin{eqnarray}
{\dot{a} (t) \over a (t)}\, \dot{h} (t) & = & 8 \pi G  \rho (t) \, \delta (t) 
\nonumber \\
\ddot{h} (t) + 3 \, {\dot{a} (t) \over a (t)}\, \dot{h} (t) 
& = & - 24 \pi \, G  \, w \, \rho(t) \delta (t)
\end{eqnarray}
with $\delta (t) = \delta \rho (t) / \rho (t) $ and $w=0$ for non-relativistic matter, 
yielding then a single equation for the trace of the metric perturbation $h(t)$,
\begin{equation}
\ddot{h} (t) + 2 \, {\dot{a} (t) \over a (t)} \dot{h} (t) \; = \; 
 - 8 \pi G ( 1 + 3\, w ) \rho(t) \delta (t) \; .
\end{equation}
Combined with the first order energy conservation 
$ - {1 \over 2}\, \left( 1 + w \right)\, h (t) \; = \; \delta (t) $, this then gives
a single equation for the density contrast $\delta(t)$,
\begin{equation}
\ddot{\delta} (t) + 2 \, {\dot{a} \over a} \, \dot{\delta} (t) - 4 \pi \, G \, \rho(t) \, \delta(t) = 0 \; .
\end{equation}
In the case of a running $G(\Box)$ these equations need to be re-derived
from the effective covariant field equations, and lead to several
additional terms not present at the classical level [23].
In other words, the correct field equations for a running $G$ are not given simply by
a naive replacement $G \rightarrow G(t) $, which would lead to incorrect results,
and violate general covariance.

It is common practice at this point to write an equation for the density contrast 
$\delta(a)$ as a function not of $t$, but of the scale factor $a(t)$, by utilizing the identities
\begin{equation}
\dot{f}(t) = a \, H \, {\partial f (a) \over \partial a}
\end{equation}
\begin{equation}
\ddot{f} (t) = {a}^2 \, {H}^2 \left( {\partial \ln H \over \partial a} + {1 \over a} \right) \,{\partial f (a) \over \partial a} + {a}^2 \, {H}^2 \, {{\partial}^2 f (a) \over \partial {a}^2}
\end{equation}
where $ f $ is any function of $ t $, and $ H \equiv \dot{a} (t) / a (t) $ is the Hubble constant.
This last quantity can then be obtained from the zero-th order $tt$ field equation
\begin{equation}
3 \, \left( { \dot{a} \over a } \right)^2 = 8 \, \pi \, G_0 \, \rho + \lambda
\end{equation}
re-written in terms of $H(a)$ as
\begin{equation}
H^2 (a) \equiv \left( {\dot{a} \over a} \right)^2 
= \left( \dot{z} \over 1 + z \right)^2 
= H_0^2 \left[ \Omega \, \left( 1 + z \right)^3 + \Omega_R \, \left( 1 + z \right)^2 + \Omega_{\lambda} \right]
\end{equation}
with $ a = {1 \over 1 + z} $ where $ z $ is a red shift, $ H_0 $ the Hubble constant evaluated today, 
and $ \Omega $ the matter (baryonic and dark) density, $ \Omega_R $ the space curvature
contribution corresponding to a curvature $ k $ term, and $\Omega_{\lambda} $the dark energy part,
\begin{equation}
\Omega_{\lambda} \equiv {\lambda \over 3 \, H^2}
\;\;\;\;\;\;
\Omega \equiv { 8 \, \pi \, G_0 \, \rho \over 3 \, H^2 }
\;\;\;\;\;\; {\rm with } \;\;\;\; 
\Omega + \Omega_{\lambda} = 1
\end{equation}
After introducing the parameter $\theta$ as the cosmological constant fraction
\begin{equation}
\theta \equiv \Omega_{\lambda} \, \left({8 \, \pi \, G \, \rho \over 3 \, H^2}\right)^{-1} 
\; = \; { \Omega_{\lambda} \over \Omega} 
\; = \; { 1 - \Omega \over \Omega } 
\end{equation}
one then obtains an equation for the density contrast $\delta (a)$ in the normal
 (i.e. non-running $G$) case
\begin{equation}
{\partial^2 \delta(a) \over \partial a^2 } + 
\left[ {\partial \ln H(a) \over \partial a} + 
{3 \over a } \right] \, {\partial \delta(a) \over \partial a} - 
4\, \pi \, G_0 \,{1 \over a^2 H(a)^2}\, \rho (a) \, \delta(a) = 0
\end{equation}
with growing solution 
\begin{equation}
\delta_0 (a) \; \sim \;  a \cdot {} _2F_1 \left({1 \over 3}, 1; {11 \over 6}; - \, a^3 \, \theta \right)
\end{equation}
where $ _2F_1 $ is a hypergeometric function.

To determine the quantum correction to $\delta (a)$ originating from 
$G(\Box)$ in Eq.~(\ref{eq:grun-box}),  one sets
\begin{equation}
\delta(a) \; = \; \delta_0 (a) \left [  1 + c_0 \, f(a) \right ]  \; ,
\end{equation}
and then uses this linear Ansatz to find the form of $ f(a) $ to lowest order in $c_0$.
The correction is unambiguously determined from the field equations with a running
Netwon constant $ G(\Box) $, but here the running of $G$ (due to the choice of variables)
naturally takes on the form
\begin{equation}
G(a) = G_0 \left [ 1 + c_a \, \left( {a \over a_{0} } \right)^{\gamma} + \dots \right ]
\label{eq:grun-a}
\end{equation}
with a scale factor $a \approx a_0$ corresponding to a mode for which $k \approx \pi / \xi$
(thus $a_0$ is not necessarily identified with the scale factor "today").
Furthermore one has $ \gamma = 3 /  2 \, \nu  $ with $ \nu = 1 / 3 $, as
determined from lattice gravity in four dimensions. 

What then remains to be done is to compute the growth index 
$ f \equiv  \partial \ln \delta  /  \partial \ln a  $, and from it
the growth index exponent $ \gamma $ defined  through 
$ f = \Omega^{\gamma} $ [24].
Ultimately one is interested in the value for this quantities in the vicinity
of a current matter fraction $\Omega \approx 0.25 $.
For a constant (i.e. not scale dependent) Newton's constant one has the
well know result $ f = 0.6028 $ and exponent $ \gamma = 0.5562 $.
An explicit calculation in the presence of a running $G$
and for a matter fraction $\Omega \approx 0.25$ gives, 
to lowest linear order in the small quantum correction $c_0$ [25],
\begin{equation}
\gamma \; = \;  0.5562 -  c_q  \, c_a \; .
\end{equation}
Here $c_a$ is the coefficient of the quantum correction  in the expression for $G(a)$, 
and therefore fixed by the underlying lattice gravity calculations, and
$c_q$ an explicitly calculable numerical constant that comes out of the solution 
of the full effective covariant field equations for $\delta(a)$.

The perturbed RW metric is well suited for discussing matter perturbations, but
occasionally one finds it more convenient to use a different metric parametrization,
 such as the one derived from the conformal Newtonian (cN) gauge line element
\begin{equation}
d \tau^2 \; = \; a^2 (t) \left \{  (1+2 \, \psi ) \, dt^2 \; - \; (1-2 \, \phi ) \, \delta_{ij} \, dx^i dx^j \right \}
\label{eq:cn-gauge}
\end{equation}
with Conformal Newtonian potentials $\psi (\vec{x},t)$ and $\phi(\vec{x},t)$.
In this gauge, and in the absence of a $G(\Box)$, the unperturbed equations are
\begin{eqnarray}
\left (  { \dot{a} \over a } \right )^2  & = &  { 8 \pi \over 3 } \, G \, a^2 \, \bar{\rho}
\nonumber \\
{ d \over d t }  \left (  { \dot{a} \over a } \right  ) & = &
- { 4 \pi \over 3 } \, G \, a^2 \, ( \bar{\rho} + 3 \bar{p} ) \; ,
\label{eq:cn_field_zeroth}
\end{eqnarray}
in the absence of spatial curvature ($k=0$).
In the presence of a running $G$ these again need to be modified, in accordance
with Eqs.~(\ref{eq:fried_rr}),  (\ref{eq:fried_tt}) and  (\ref{eq:grun_frw}).
A cosmological constant can be conveniently included in the $\bar{\rho}$ and $\bar{p}$, 
with $\bar{\rho}_\lambda = \lambda/ 8 \pi G = - \bar{p}_\lambda $.
In this gauge scalar perturbations are characterized by Fourier modes 
$\psi(\vec{k},t)$ and $\phi (\vec{k},t)$, and the first order Einstein field equations
in the absence of $G(\Box)$ read [26]
\begin{eqnarray}
k^2 \, \phi \, + \, 3 \, { \dot{a} \over a }  \, \left ( \dot{\phi} \, + \, { \dot{a} \over a }  \, \psi \right )
& = &
4 \pi \, G \, a^2  \,  \delta T^0_{\;\; 0}
\nonumber \\
k^2  \, \left ( \dot{\phi} \, + \, { \dot{a} \over a }  \, \psi \right )
& = &
4 \pi \, G \, a^2  \, ( \bar{\rho} + \bar{p} ) \, \theta
\nonumber \\
\ddot{\phi} \, + \, { \dot{a} \over a }  \left ( 2 \dot{\phi} \, + \dot{\psi} \right )
+ \left ( 2 \, { \ddot{a} \over a }  \, - \, { \dot{a}^2 \over a^2 }   \right ) \, \psi
\, + \, { k^2 \over 3 } \, ( \phi \, - \, \psi )
& = &
{ 4 \pi \over 3 } \, G \, a^2  \,  \delta T^i_{\;\; i}
\nonumber \\
k^2 \, ( \phi \, - \, \psi )
& = &
12 \pi \, G \, a^2  \,  ( \bar{\rho} + \bar{p} ) \, \sigma \;\;\;\;
\label{eq:cn_field_pert}
\end{eqnarray}
where the perfect fluid energy-momentum tensor is given to linear order in the 
perturbations  
$\delta \rho = \rho - \bar{\rho} $ and $\delta p = p - \bar{p} $
by
\begin{eqnarray}
T^0_{\;\; 0}  & = &  - ( \bar{\rho} \, + \, \delta \rho )
\nonumber \\
T^0_{\;\; i}  & = &  ( \bar{\rho} \, + \, \bar{p} ) \, v_i  \; = \; - T^i_{\;\; 0}
\nonumber \\
T^i_{\; j}  & = &  ( \bar{p} \, + \, \delta p ) \, \delta^i_{\; j} \, + \, \Sigma^i_{\; j} 
\;\;\;\; \Sigma^i_{\; i}=0
\end{eqnarray}
and one has allowed for an anisotropic shear perturbation $\Sigma^i_{\; j}$ to
the perfect fluid form $T^i_{\; j}$.
 The two quantities $\theta$ and $\sigma$ are commonly defined by 
\begin{equation}
( \bar{\rho} \, + \, \bar{p} )  \, \theta \; \equiv \; i \, k^j \, \delta T^0_{\; j}  
\;\;\;\;  
( \bar{\rho} \, + \, \bar{p} )  \, \sigma \; \equiv \; 
- ( \hat{k_i} \hat{k_j} - { 1 \over 3} \delta_{ij} ) \Sigma^i_{\; j}
\end{equation}
with $\Sigma^i_{\; j} \equiv T^i_{\, j} - \delta^i_{\; j} T^k_{\; k} /3 $ the traceless
component of $T^i_{\; j}$.
For a perfect fluid $\theta$ is the divergence of the fluid velocity, $\theta = i k^j v_j$, 
with $v^j = d x^j / dt $ the small velocity of the fluid.
The field equations imply, by consistency, the covariant energy momentum conservation law
\begin{eqnarray}
\dot{\delta} & = & - (1+w) \, (\theta - 3 \dot{\phi} ) - 3 \, { \dot{a} \over a }  \, 
\left ( { \delta p \over \delta \rho } - w \right ) \delta
\nonumber \\
\dot{\theta} & = &  - { \dot{a} \over a }  \, ( 1 - 3 w ) \, \theta - 
{ \dot{w} \over 1+w } \, \theta 
+ { 1 \over 1+ w } { \delta p \over \delta \rho }  \, k^2 \delta - k^2 \sigma + k^2 \psi 
\end{eqnarray}
and relate the matter fields $\delta$, $\sigma$ and $\theta$ to the metric perturbations
$\phi$ and $\psi$.
where $\delta $ is the matter density contrast $\delta = \delta \rho / \rho $, and
$w$ is the equation of state parameter $w= p/\rho $.

In the presence of a $G(\Box)$ the above equations need to be re-derived and amended [25],
starting from the covariant field equations of Eq.~(\ref{eq:field1}) in the cN gauge
of Eq.~(\ref{eq:cn-gauge}),  with zero-th order modified field equations as
in Eqs.~(\ref{eq:fried_tt}) and (\ref{eq:fried_rr}), using the expansion
for $G(\Box)$ given in Eq.~(\ref{eq:gbox_h}), but now in terms of the new cN
gauge potentials $\phi$ and $\psi$.
One key question is then the nature of the vacuum-polarization induced 
anisotropic shear perturbation correction $\Sigma^i_{\; j}$ 
appearing in the covariant effective field equations analogous to 
Eqs.~(\ref{eq:cn_field_pert}),
but derived with a $G(\Box)$.
In particular  one would expect the quantum correction to the energy momentum
tensor appearing on the r.h.s. of Eq.~(\ref{eq:field1}) to contribute new terms
to the last of Eqs.~(\ref{eq:cn_field_pert}),
which could then account for a non-zero stress $\sigma$, and thus for a small 
deviation from the classical result for a perfect fluid, $\phi = \psi$.

\vskip 20pt

{\bf Acknowledgements}

I wish to thank the organizers of the MG-12 conference, and in particular
Thibault d'Amour, Gabriele Veneziano and Remo Ruffini, for the opportunity
of attending the conference, 
and numerous discussions relevant to the work presented here.
I have also benefitted from conversations with Alexey Vikhlinin regarding
the past and future observational constraints on the density contrast exponents from
 the study of large galactic clusters.
I furthermore wish to express my gratitude to my collaborator Ruth M. Williams and 
my student Reiko Toriumi, for contributing to the work described here.
Finally I thank Hermann Nicolai and the
Max Planck Institut f\" ur Gravitationsphysik (Albert-Einstein-Institut)
in Berlin for hospitality when part of the work
presented in this review was performed, and to the NSF for financial support.


\vfill
\newpage

\end{document}